\documentclass[11pt,a4paper]{article}

\usepackage{amstext,amsgen,latexsym}
\usepackage{amstext,amssymb,amsfonts,latexsym}
\usepackage{theorem}
\usepackage{pifont}

 \newcommand{\bs}{\bigskip}
 \newcommand{\ms}{\medskip}
 \newcommand{\n}{\noindent}
 \newcommand{\s}{\smallskip}
 \newcommand{\hs}[1]{\hspace*{ #1 mm}}
 \newcommand{\vs}[1]{\vspace*{ #1 mm}}

 \newcommand{\real}{\mathbb{R}}

 \newcommand{\nat}{\mathbb{N}}

 \newcommand{\ie}{\textrm{i.e.},\hspace*{2mm}}
 
 \newcommand{\etal}{\textrm{et al.}\hspace*{2mm}}
 \newcommand{\etalc}{\textrm{et al.}}

 \newcommand{\AAA}{{\cal A}}

 \newcommand{\CC}{{\cal C}}
 
 \newcommand{\FF}{{\cal F}}

 \newcommand{\GG}{{\cal G}}

 \newcommand{\UU}{{\cal U}}

 \def\bbox{\vrule height6pt width6pt depth1pt}

\theoremstyle{plain}
\theoremheaderfont{\bfseries}
 \newtheorem{theorem}{Theorem}[section]
 \newtheorem{lemma}[theorem]{Lemma}
 \newtheorem{proposition}[theorem]{Proposition}

 {\theorembodyfont{\rmfamily}
  }
 {\theorembodyfont{\rmfamily} }
 {\theorembodyfont{\rmfamily} }

 \newenvironment{proofsketch}{\par \noindent
            {\bf Proof Sketch. \hs{2}}}{\hfill\bbox \vspace*{3mm}}

 \newenvironment{proofof}[1]{\vspace*{5mm} \par \noindent
         {\bf Proof of #1.\hs{2}}}{\hfill$\Box$ \vspace*{3mm}}
         
 \newenvironment{proofsketchof}[1]{\vspace*{5mm} \par \noindent
         {\bf Proof Sketch of #1.\hs{2}}}{\hfill$\Box$ \vspace*{3mm}}

 \newcommand{\pair}[1]{\langle #1 \rangle}

\newif\ifnotesw\noteswtrue
   {\ifnotesw\marginpar[\hfill\(\top\)]{\(\top\)}\fi}%
      {\ifnotesw\marginpar[\hfill\(\bot\)]{\(\bot\)}\fi}
      
\newcommand{\mnote}[1]%
   {\ifnotesw\marginpar%
	  [{\scriptsize\begin{minipage}[t]{\marginparwidth}
	  \raggedleft#1%
		  \end{minipage}}]%
	  {\scriptsize\begin{minipage}[t]{\marginparwidth}
	  \raggedright#1%
		  \end{minipage}}%
    \fi}

\newcommand{\ignore}[1]{}

\newcommand{\maxcsp}{\mathrm{MAX\mbox{-}CSP}}

\newcommand{\maxprodcsp}{\mathrm{MAX\mbox{-}PROD\mbox{-}CSP}}
\newcommand{\maxprodcspstar}{\mathrm{MAX\mbox{-}PROD\mbox{-}CSP}^{*}}

\newcommand{\apx}{\mathrm{APX}}
\newcommand{\expapx}{\mathrm{exp\mbox{-}APX}}
\newcommand{\po}{\mathrm{PO}}
\newcommand{\npo}{\mathrm{NPO}}

\newcommand{\sol}{\mathrm{sol}}

\newcommand{\APTreduces}{\leq_{\mathrm{APT}}}
\newcommand{\APTequiv}{\equiv_{\mathrm{APT}}}
\newcommand{\AF}{{\cal AF}}
\newcommand{\DG}{{\cal DG}}
\newcommand{\NZ}{{\cal NZ}}
\newcommand{\ED}{{\cal ED}}
\newcommand{\IM}{{\cal IM}}
\newcommand{\IMopt}{{\cal IM}_{opt}}

\begin{document}
\pagestyle{plain}
\setcounter{page}{1}

\begin{center}
{\Large {\bf Optimization, Randomized Approximability, and \s\\
Boolean Constraint Satisfaction Problems}} \bs\bs\\

{Tomoyuki Yamakami}\footnote{Present Affiliation: Department of Information Science, University of Fukui, 3-9-1 Bunkyo, Fukui 910-8507,  Japan} \bs\\
\end{center}

\begin{quote}
{\bf Abstract.}\hs{1}
We give a unified treatment to optimization problems that can be 
expressed in the form of nonnegative-real-weighted Boolean constraint satisfaction problems. Creignou, Khanna, Sudan, Trevisan, and Williamson  studied the complexity of approximating their optimal solutions whose optimality is measured by the sums of outcomes of constraints. 
To explore a wider range of  optimization constraint satisfaction problems, 
following an early work of Marchetti-Spaccamela and Romano, we study the case where the optimality is measured by products of constraints' outcomes. We completely classify those problems into three categories:  PO problems, NPO-hard problems, and intermediate problems that lie between the former two categories. 
To prove this trichotomy theorem, we analyze characteristics of nonnegative-real-weighted constraints using a variant of the notion of 
T-constructibility developed earlier for complex-weighted counting constraint satisfaction problems.  

\ms

\n{\bf keywords:} optimization problem, approximation algorithm, 
constraint satisfaction problem, PO, APX, approximation-preserving reducibility
\end{quote} 

\section{Maximization by Multiplicative Measure}\label{sec:introduction}

In the 1980s started extensive studies that have greatly improved our  understandings of the exotic behaviors of  
various optimization problems within a scope of computational complexity theory. These studies have brought us deep insights into the approximability and inapproximability of optimization problems; however, many studies have targeted individual problems by cultivating 
different and independent methods for them.  To push our insights deeper, we are   focused on a collection of ``unified'' optimization problems,  
whose foundations are all formed in terms of {\em Boolean constraint satisfaction problems} (or {\em CSPs}, in short). Creignou is the first to have given a formal treatment to maximization problems derived from CSPs \cite{Cre95}.  
The {\em maximization constraint satisfaction problems} (or MAX-CSPs for succinctness) are, in general, optimization problems in which we seek a truth assignment $\sigma$ of Boolean variables that maximizes an {\em objective value}\footnote{A function that associates an objective value (or a measure) to each solution is called  an {\em objective function} or (a {\em measure function}).} (or a {\em measure}) of $\sigma$, which equals the number of constraints being satisfied at once. 
Creignou presented three criteria (which are $0$-validity, $1$-validity, and $2$-monotonicity) under which we can solve the MAX-CSPs in polynomial time;  that is, the problems belong to $\po$. 

Creignou's result was later reinforced by Khanna, Sudan, Trevisan, and Williamson \cite{KSTW01}, who gave a unified treatment to several types of CSP-based optimization problems, including $\mathrm{MAX\mbox{-}CSP}$, $\mathrm{MIN\mbox{-}CSP}$, and $\mathrm{MAX\mbox{-}ONE\mbox{-}CSP}$. 
With constraints limited to ``nonnegative'' integer values, Khanna et al. defined    $\maxcsp(\FF)$ as the maximization problem in which constraints are all taken from constraint set $\FF$ and the maximization is measured by the ``sum'' of the objective values of constraints. For a later comparison, we call such a measure an {\em additive measure}. 
More formally, $\maxcsp(\FF)$ is defined as:

\ms
$\maxcsp(\FF)$:
\begin{itemize}\vs{-2}
\item {\sc Instance:} a {\em finite} set $H$ of elements of the form $\pair{h,(x_{i_1},\ldots,x_{i_k})}$ on Boolean variables $x_1,\ldots,x_n$, where $h\in\FF$, $\{i_1,\ldots,i_k\}\subseteq[n]$, 
and $k$ is the arity of $h$.
\vs{-2}
\item {\sc Solution:} a truth assignment $\sigma$ to the variables  $x_1,\ldots,x_n$.   
\vs{-2}
\item {\sc Measure:}  the sum $\sum_{\pair{h,x'}\in H}h(\sigma(x_{i_1}),\ldots,\sigma(x_{i_k}))$, where $x'=(x_{i_1},\ldots,x_{i_k})$. 
\end{itemize}
\vs{-1}

\sloppy For instance, $\maxcsp(XOR)$ coincides with the optimization problem $\mathrm{MAX\mbox{-}CUT}$, which is known to be $\mathrm{MAX\mbox{-}SNP}$-complete \cite{PY91}. 
Khanna et al. re-proved Creignou's classification theorem that, for every set $\FF$ of constraints, if $\FF$ is one of $0$-valid, $1$-valid, and $2$-monotone, then $\maxcsp(\FF)$ is in $\po$, and otherwise, $\maxcsp(\FF)$ is $\apx$-complete
\footnote{$\apx$ is the collection of optimization problems whose 
optimal solutions can be (deterministically) approximated to within 
a fixed constant in polynomial time.} 
under their approximation-preserving reducibility. 
This classification theorem was proven by an extensive use of a notion of {\em strict/perfect implementation}. 

{}From a different and meaningful perspective, Marchetti-Spaccamela and Romano \cite{MR85} made a general discussion on max NPSP problems whose maximization is measured by the ``products'' of the objective values of chosen  (feasible) solutions. {}From their general theorem follows 
an early result of \cite{AMP81} that the maximization problem, MAX-PROD-KNAPSACK, has a {\em fully polynomial(-time) approximation scheme}  (or FPTAS). This result can be compared with another result of Ibarra and Kim \cite{IK75} that MAX-KNAPSACK (with additive measures) admits an FPTAS. 
The general theorem of \cite{MR85} requires development of a new proof technique, called {\em variable partitioning}. 
A similar approach was taken in a study of linear multiplicative programming to minimize the ``product'' of two positive linear cost functions, subject to linear constraints \cite{KK92}.  
In contrast with the previous additive measures, we call this different   type of measures {\em multiplicative measures}, and we wish to study the behaviors of MAX-CSPs whose maximization is taken over multiplicative measures. 
 
Our approach clearly fills a part of what Creignou \cite{Cre95} and Khanna \etalc~\cite{KSTW01} left unexplored. Let us formalize our objectives for clarity. To differentiate our multiplicative measures from additive measures, we develop 
a new notation $\maxprodcsp(\cdot)$, in which ``PROD'' refers to  a use of  ``product'' of objective values. 

\ms
$\maxprodcsp(\FF)$:
\begin{itemize}\vs{-2}
\item {\sc Instance:} a {\em finite} set $H$ of elements of the form $\pair{h,(x_{i_1},\ldots,x_{i_k})}$ on Boolean variables $x_1,\ldots,x_n$, where $h\in\FF$ and $\{i_1,\ldots,i_k\}\subseteq[n]$.  
\vs{-2}
\item {\sc Solution:} a truth assignment $\sigma$ to $x_1,\ldots,x_n$.   
\vs{-2}
\item {\sc Measure:}  the product $\prod_{\pair{h,x'}\in H}h(\sigma(x_{i_1}),\ldots,\sigma(x_{i_k}))$, where $x'=(x_{i_1},\ldots,x_{i_k})$. 
\end{itemize}
\vs{-1}

There is a natural, straightforward way to relate $\maxcsp(\FF)$'s to 
$\maxprodcsp(\FF)$'s. For example, consider the problem MAX-CUT and 
its multiplicatively-measured counterpart, MAX-PROD-CUT. 
Given any cut for MAX-CUT, we set the weight of each vertex to be $2$ if it belongs to the cut, and set the weight to be $1$ otherwise. When the cardinality of a  cut is maximized, the same cut incurs the maximum product value as well. 
In other words, an algorithm that ``exactly'' finds an optimal solution for MAX-CUT also computes an optimal solution for MAX-PROD-CUT. However, when an algorithm only tries to ``approximate'' such an optimal solution, its performance rates for MAX-CUT and for MAX-PROD-CUT significantly differ. 
In a case of geometric programming, a certain type of product objective function is known to be ``reduced'' to additive ones if function values are all positive (see an survey \cite{BKVH07}). In our setting, a different  complication comes in when some values of $h$'s accidentally fall into zero and thus the entire product value vanishes.  
Thus, an approximation algorithm using an additive measure does not seem to  lead to an approximation algorithm with a multiplicative measure.   
This circumstance indicates that multiplicative measures appear to endow their associated optimization problems  with much higher approximation complexity than additive measures do. We then need to develop a quite different methodology for a study on the approximation complexity of multiplicatively-measured optimization problems. 

Within the framework of MAX-PROD-CSPs, we can precisely  
express many maximization problems, such as 
MAX-PROD-CUT, MAX-PROD-SAT, MAX-PROD-IS (independent set), and MAX-PROD-BIS (bipartite independent set), which are naturally induced from 
their corresponding additively-measured maximization problems. 
Some of their formal definitions will be given in Section \ref{sec:optimization}. 
In a context of approximability, since the multiplicative measures can provide  a richer structure than the additive measures, classification theorems for MAX-CSPs and for MAX-PROD-CSPs are essentially different.   
The classification theorem for MAX-PROD-CSPs---our main result---is formally 
stated in Theorem \ref{main-theorem}, in which  we use an abbreviation, $\maxprodcspstar(\FF)$, 
to mean a MAX-PROD-CSP whose unary constraints are provided for 
free\footnote{Allowing a free use of arbitrary unary constraints is a commonly used assumption for decision CSPs and counting CSPs.} 
and other nonnegative-real-weighted constraints are drawn from $\FF$.   
\begin{theorem}\label{main-theorem}
Let $\FF$ be any set of constraints. If either $\FF\subseteq \AF$ or $\FF\subseteq\ED$, then $\maxprodcspstar(\FF)$ is in $\po$. Otherwise, if $\FF\subseteq \IMopt$, then $\maxprodcspstar(\FF)$ is APT-reduced from $\mathrm{MAX\mbox{-}PROD\mbox{-}BIS}$ and is APT-reduced to $\mathrm{MAX\mbox{-}PROD\mbox{-}FLOW}$. Otherwise, $\maxprodcspstar(\FF)$ is APT-reduced  from $\mathrm{MAX\mbox{-}PROD\mbox{-}IS}$. 
\end{theorem}
Here, ``APT'' stands for ``approximation preserving Turing'' in the sence of \cite{DGGJ03}. Moreover,  $\AF$ is the set of  ``affine''-like constraints, $\ED$ is related to the binary equality and disequality constraints, and $\IMopt$ is characterized by ``implication''-like constraints. The problem MAX-PROD-FLOW is a maximization problem of finding a design that maximizes the amount of water flow in a given direct graph. See Sections \ref{sec:constraint}--\ref{sec:optimization} for their formal definitions.

The purpose of the rest of this paper is to prove Theorem \ref{main-theorem}. For this purpose, we will introduce a new notion of {\em T$_{max}$-constructibility}, which is a variant of the notion of T-constructibility invented in \cite{Yam10a}. This T$_{\mathrm{max}}$-constructibility is proven to be a powerful tool in dealing with MAX-PROD-CSPs. 

\section{Formal Definitions and Basic Properties}\label{sec:basic-definition}

Let $\nat$ denote the set of all {\em natural numbers} (\ie nonnegative integers) and let $\real$ denote the set of all real numbers.  
For convenience, $\nat^{+}$ expresses $\nat-\{0\}$ and, for each 
$n\in\nat^{+}$, $[n]$ denotes the integer set $\{1,2,\ldots,n\}$. 
Moreover, the notation $\real^{\geq0}$ stands for the set $\{r\in\real\mid r\geq0\}$. 

\subsection{Constraints and Relations}\label{sec:constraint}

Here, we use terminology given in \cite{Yam10a}. 
A {\em (nonnegative-real-weighted) constraint} is a function mapping from $\{0,1\}^k$ to $\real^{\geq0}$, where 
$k$ is the {\em arity} of $f$.  
Assuming the standard lexicographic order on $\{0,1\}^k$, we express  $f$   as a series of its output values. 
For instance, if $k=2$, then $f$ is $(f(00),f(01),f(10),f(11))$. 
We set $EQ=(1,0,0,1)$, $\Delta_0=(1,0)$, and $\Delta_1=(0,1)$.  

A {\em relation} of arity $k$ is a subset of $\{0,1\}^k$. Such a relation can be also viewed as a function mapping Boolean variables to $\{0,1\}$ (\ie $x\in R$ iff $R(x)=1$, for every $x\in\{0,1\}^k$) and it can be treated as a Boolean constraint. For instance, logical relations $OR$, $NAND$, $XOR$, and $Implies$ are all expressed as appropriate constraints in the following manner: 
$OR=(0,1,1,1)$, $NAND=(1,1,1,0)$, $XOR=(0,1,1,0)$, and $Implies=(1,1,0,1)$. 
A relation $R$ is {\em affine} if it is expressed as a set of solutions to a certain system of linear equations over $GF(2)$. An {\em underlying relation} $R_f$ of $f$ is defined as $R_f=\{x\mid f(x)\neq0\}$. 

We introduce the following six special sets of constraints. 

\begin{enumerate}\vs{-1}
\item Let $\UU$ denote the set of all unary constraints. 
\vs{-2}
\item The notation $\NZ$ expresses the set of all non-zero constraints.
\vs{-2}
\item Denote by $\DG$ the set of all constraints that are expressed by products of unary constraints, each of which is applied to a different variable. Such a constraint is called {\em degenerate}. 
\vs{-2}
\item Let $\ED$ 
denote the set of all constraints that are expressed as products of some of 
unary constraints, the equality $EQ$, and the disequality $XOR$. 
\vs{-2}
\item The set $\AF$ is defined as the collection of all constraints of the form 
$g(x_1,\ldots,x_k) \prod_{j:j\neq i}R_j(x_i,x_j)$ for a certain fixed index $i\in[k]$, where  $g$ is in $\DG$ and each $R_j$ is an affine relation. 
\vs{-2}
\item Define $\IMopt$ to be the collection of all constraints that are products of some of the following constraints: unary constraints and constraints of the form $(1,1,\lambda,1)$ with $0\leq \lambda<1$.  This is  different from $\IM$ defined in \cite{Yam10a}.
\end{enumerate}

\begin{lemma}\label{xw-yz-member-IMopt}
For any constraint $f=(x,y,z,w)$ with $x,y,z,w\in\real^{\geq0}$, if $xw>yz$, then $f$ belongs to $\IMopt$. 
\end{lemma}

\subsection{Optimization Problems with Multiplicative Measures}\label{sec:optimization}

A {\em  (combinatorial) optimization problem} $P=(I,\sol,m)$ 
takes input instances of 
 ``admissible data'' to the target problem. We often write $I$ to denote the set of all such instances and $\sol(x)$ denotes a set of {\em (feasible) solutions} associated with instance $x$. A {\em measure function} (or {\em objective function}) $m$ associates a nonnegative real number to each solution $y$ in $\sol(x)$; that is, $m(x,y)$ is an 
{\em objective value} (or a {\em measure}) of the solution $y$ on the instance $x$.  
We conveniently assume $m(x,y)=0$ for any element $y\not\in\sol(x)$. 
The goal of the problem $P$  is to find a solution $y$ in $\sol(x)$ that has an optimum value (such $y$ is called an {\em optimal solution}), where the optimality is measured by either the maximization or minimization of an objective value $m(x,y)$, taken over all solutions $y\in\sol(x)$. When $y$ is an optimal solution, we set $m^{*}(x)$ to be $m(x,y)$. 
 
Let $\npo$ denote the class of all optimization problems $P$ such that (1) input instances and solutions can be recognized in polynomial time; (2) solutions are polynomially-bounded in input size; and (3) a measure function can be computed in polynomial time.    
Define $\po$ as the class of all problems $P$ in $\npo$ such that there exists a deterministic algorithm that, for every instance $x\in I$, returns an optimal solution $y$ in $\sol(x)$ in time polynomial in the size $|x|$ of the instance $x$.   

We say that, for a fixed real-valued function $\alpha$ with $\alpha(n)\geq1$ for any input size $n\in\nat$, an algorithm $\AAA$ for an optimization problem $P=(I,\sol,m)$ is an {\em $\alpha$-approximation algorithm} if, for every instance $x\in I$, $\AAA$ produces a solution $y\in\sol(x)$ satisfying that  $1/\alpha(|x|)\leq \left|  {m(x,y)}/{m^{*}(x)} \right|\leq \alpha(|x|)$, except that, whenever $m^*(x)=0$, we always demand that $m(x,y)=0$.  
Such a $y$ is called an {\em $\alpha(n)$-approximate solution} 
for input instance $x$ of size $n$.   
The class $\apx$ (resp., $\expapx$) consists of all problems $P$ in $\npo$ such that there are a constant $r\geq1$ (resp., an exponentially-bounded\footnote{This function $\alpha$ must satisfy that there exists a positive polynomial $p$ for which  $1\leq\alpha(n)\leq2^{p(n)}$ for any number $n\in\nat$.} function $\alpha$) and a polynomial-time $r$-approximation (resp., $\alpha$-approximation) algorithm for $P$.
Approximation algorithms are often randomized.  
A {\em randomized approximation scheme} for $P$  is a randomized algorithm that takes a standard input instance $x\in I$ together with an error tolerance parameter $\varepsilon\in(0,1)$, and outputs a $2^{\varepsilon}$-approximate solution $y\in \sol(x)$ with probability at least $3/4$. 

Of numerous existing notions of approximation-preserving reducibilities, we choose a notion introduced recently by Dyer, Goldberg, Greenhill, and Jerrum \cite{DGGJ03}, which can be viewed as a randomized variant of Turing reducibility, based on a mechanism of {\em oracle Turing machine}. 
Since the purpose of Dyer \etal is to solve counting problems approximately, we need to modify their notion so that we can deal with optimization problems.  
Given two optimization problems $P=(I,\sol,m)$ and $Q=(I',\sol',m')$, 
a {\em polynomial-time  (randomized) approximation-preserving Turing reduction} (or {\em APT-reduction}, in short) from $P$ to $Q$ is a randomized algorithm $N$ that takes a pair $(x,\varepsilon)\in I \times(0,1)$ as input,  
uses an arbitrary randomized approximation scheme (not necessarily polynomial time-bounded) $M$ for $Q$ as {\em oracle}, 
and satisfies the following conditions:
(i) $N$ is a randomized approximation scheme for $P$ for any choice of oracle $M$ for $Q$;  
(ii) every {\em oracle call} made by $N$ is of the form $(w,\delta)\in I' \times(0,1)$ satisfying  
$1/\delta \leq p(|x|,1/\varepsilon)$, where $p$ 
is a certain absolute polynomial, 
and an oracle answer is an outcome of $M$ on 
the input $(w,\delta)$; and 
(iii) the running time of $N$ is bounded from above by a certain polynomial in $(|x|,1/\varepsilon)$, not depending on the choice of the oracle $M$. In this case, we write $P\APTreduces Q$ and we also say that $P$ is {\em APT-reducible} (or {\em APT-reduced}) to $Q$. Note that APT-reducibility composes. If $P\APTreduces Q$ and $Q\APTreduces P$, then $P$ and $Q$ 
are said to be {\em APT-equivalent} and we use the notation $P\APTequiv Q$. 


In the definition of $\maxprodcsp(\FF)$ given in Section \ref{sec:introduction}, 
we also write $h(x_{i_1},\ldots,x_{i_k})$ to mean  $\pair{h,(x_{i_1},\ldots,x_{i_k})}$ in $H$. For notational simplicity, 
we intend to write, for example, $\maxprodcsp(f,\FF,\GG)$ instead of  $\maxprodcsp(\{f\}\cup\FF\cup\GG)$. In addition, we abbreviate as  $\maxprodcspstar(\FF)$ the maximization problem $\maxprodcsp(\FF\cup\UU)$. 

For any optimization problem $P$ and any class $\CC$ of optimization problems, we write $P\APTreduces \CC$ if there exists a problem $Q\in\CC$ such that $P\APTreduces Q$. Our choice of APT-reducibility makes it possible to prove Lemma \ref{exp-APX-bound}; however,  it is not clear whether the lemma implies that $\maxprodcsp(\FF)\in \expapx$.

\begin{lemma}\label{exp-APX-bound}
 $\maxprodcsp(\FF)\APTreduces \expapx$ for any constraint set $\FF$. 
\end{lemma}

Hereafter, we introduce the maximization problems stated in Theorem \ref{main-theorem}. 

\ms

MAX-PROD-IS: (maximum product independent set)
\begin{itemize}\vs{-2}
\item {\sc Instance:} an undirected graph $G=(V,E)$ and a series $\{w_x\}_{x\in V}$ of vertex weights with $w_x\in\real^{\geq0}$; 
\vs{-2}
\item {\sc Solution:} an independent set $A$ on $G$; 
\vs{-2}
\item {\sc Measure:} the product $\prod_{x\in A}w_x$.
\end{itemize}

This maximization problem MAX-PROD-IS literally coincides with    $\maxprodcsp(NAND,\GG_0)$, where $\GG_0=\{[1,\lambda]\mid \lambda\geq0\}$, 
and it can be easily shown to be $\npo$-complete. 
When all input graphs are limited to bipartite graphs, 
the corresponding problem is called MAX-PROD-BIS.

\ms

MAX-PROD-BIS: (maximum product bipartite independent set)
\begin{itemize}\vs{-2}
\item In MAX-PROD-IS, all input graphs are 
limited to bipartite graphs.
\end{itemize}

Since the above two problems can be expressed in the form of 
$\maxprodcspstar(\cdot)$, we can draw the following important conclusion, which 
becomes part of the proof of the main theorem.

\begin{lemma}\label{OR-complete}
\begin{enumerate}
\item $\mathrm{MAX\mbox{-}PROD\mbox{-}IS}\APTreduces \maxprodcspstar(OR)$.
\vs{-2}
\item $\mathrm{MAX\mbox{-}PROD\mbox{-}BIS} \APTreduces \maxprodcspstar(Implies)$.
\end{enumerate}
\end{lemma}


Next, we introduce a special maximization problem, called MAX-PROD-FLOW, whose intuitive setting is explained as follows.  
Suppose that water flows from point $u$ to point $v$ through a one-way pipe at flow rate $\rho_{(u,v)}$. A value $\sigma(x)$ expresses an elevation (indicating either the {\em bottom level} or the {\em top level}) of point $x$ so that water runs from point $u$ to point $v$ whenever  $\sigma(u)\geq\sigma(v)$. More water is added at influx rate $w_{v}$ at point $v$ for which $\sigma(v)=1$.  

\ms

MAX-PROD-FLOW: (maximum product flow)
\begin{itemize}\vs{-2}
\item {\sc Instance:} a directed graph $G=(V,E)$, a series $\{\rho_e\}_{e\in E}$ of flow rates with $\rho_e\geq1$, and a series 
$\{w_x\}_{x\in V}$ of influx rates with $w_x\geq0$; 
\vs{-2}
\item {\sc Solution:} a Boolean assignment $\sigma$ to $V$;
\vs{-2}
\item {\sc Measure:} the product $\left(\prod_{(x,y)\in E,\sigma(x)\geq\sigma(y)} \rho_{(x,y)}\right) \left(\prod_{z\in V,\sigma(z)=1} w_z\right)$.
\end{itemize}

{}From the above definition, it is not difficult to prove the following statement. 

\begin{lemma}\label{IMopt-to-FLOW}
For any constraint set $\FF\subseteq \IMopt$, $\maxprodcspstar(\FF)$ is APT-reducible to $\mathrm{MAX\mbox{-}PROD\mbox{-}FLOW}$.
\end{lemma}

\subsection{T$_{\mathrm{max}}$-Constructibility}\label{sec:T-constructibility}

\sloppy To pursue notational succinctness, we use the following notations. Let $f$ be any arity-$k$ constraint. For any two distinct indices $i,j\in[k]$ and any bit $c\in\{0,1\}$, 
let $f^{x_i=c}$ denote the function $g$ satisfying that $g(x_1,\ldots,x_{i-1},x_{i+1},\ldots,x_k) = f(x_1,\ldots,x_{i-1},c,x_{i+1},\ldots,x_k)$ and let $f^{x_i=x_j}$ be  the function $g$ defined as  $g(x_1,\ldots,x_{i-1},x_{i+1},\ldots,x_k) =
 f(x_1,\ldots,x_{i-1},x_{j},x_{i+1},\ldots,x_k)$. 
Moreover, we denote by $\max_{y_1,\ldots,y_d}(f)$ the function $g$ defined as $g(x_1,\ldots,x_k) = \max_{(y_1,\ldots,y_d)\in\{0,1\}^d}\left\{ f(x_1,\ldots,x_k,y_1,\ldots,y_d) \right\}$, where $y_1,\ldots,y_d$ are all distinct and different from $x_1,\ldots,x_k$, and let $\lambda\cdot f$ denote the function satisfying  $(\lambda \cdot f)(x_1,\ldots,x_k) = \lambda\, f(x_1,\ldots,x_k)$. 

A helpful tool invented in \cite{Yam10a} for counting CSPs  
is a notion of {\em T-constructibility}. For our purpose of proving the main theorem, 
we wish to modify this notion and introduce a notion of 
T$_{\mathrm{max}}$-constructibility. 
We say that an arity-$k$ constraint $f$ is {\em T$_{max}$-constructible} (or {\em T$_{max}$-constructed}) from a constraint set $\GG$ if $f$ can be obtained, initially from constraints in $\GG$, by applying recursively a finite number (possibly zero) of seven functional operations described below.

\begin{enumerate}\vs{-1}
\item {\sc Permutation:} for two indices $i,j\in[k]$ with $i<j$, 
by exchanging two columns $x_i$ and $x_j$ in  $(x_1,\ldots,x_i,\ldots,x_j,\ldots,x_k)$, transform $g$ into $g'$, 
where $g'$ is defined as $g'(x_1,\ldots,x_i,\ldots,x_j,\ldots,x_k) = g(x_1,\ldots,x_j,\ldots,x_i,\ldots,x_k)$.
\vs{-2}
\item {\sc Pinning:} for an index $i\in[k]$ and a bit $c\in\{0,1\}$, build $g^{x_i=c}$ from $g$.
\vs{-2}
\item {\sc Linking:} for two distinct indices $i,j\in[k]$, build $g^{x_i = x_j}$ from $g$.
\vs{-2}
\item \sloppy {\sc Expansion:} for an index $i\in[k]$, introduce a new ``free'' variable, say, $y$ and  transform $g$ into $g'$ that is defined by $g'(x_1,\ldots,x_i,y,x_{i+1},\ldots,x_k) =  g(x_1,\ldots,x_{i},x_{i+1},\ldots,x_k)$.
\vs{-2}
\item {\sc Multiplication:} from two constraints $g_1$ and $g_2$ of arity $k$ that share the same input variable series $(x_1,\ldots,x_k)$, build  the  constraint $g_1\cdot g_2$, where $(g_1\cdot g_2)(x_1,\ldots,x_k) = g_1(x_1,\ldots,x_k)g_2(x_1,\ldots,x_k)$.
\vs{-2}
\item {\sc Maximization:} build $\max_{y_1,\ldots,y_d}(g)$ from $g$, where 
 $y_1,\ldots,y_d$ are not shared with any other constraint other than this particular constraint $g$. 
\vs{-2}
\item {\sc Normalization:}  for a positive constant $\lambda$, build $\lambda\cdot g$ from $g$. 
\end{enumerate}\vs{-1}
When $f$ is T$_{\mathrm{max}}$-constructible from $\GG$, we use the notation $f\leq_{con}^{max}\GG$. In particular, when $\GG$ is a singleton $\{g\}$, we also write $f\leq_{con}^{max}g$ instead of $f\leq_{con}^{max}\{g\}$. 

It holds that T$_{\mathrm{max}}$-constructibility between constraints guarantees APT-reducibility between their corresponding $\maxprodcspstar(\cdot)$'s. 

\begin{lemma}\label{maxcon-to-APreduce}
If $f\leq_{con}^{max}\GG$, then $\maxprodcspstar(f,\FF)$ is APT-reducible to $\maxprodcspstar(\GG,\FF)$ for any constraint set $\FF$. 
\end{lemma}

\section{Proof of the Main Theorem }\label{sec:main-theorem}

Our main theorem---Theorem \ref{main-theorem}---states that all maximization problems of the form of $\maxprodcspstar(\cdot)$ can be classified into three categories. This trichotomy theorem sheds a clear contrast with  the dichotomy theorem of Khanna \etalc~\cite{KSTW01} for MAX-CSPs. 
Hereafter, we will present the proof of Theorem \ref{main-theorem}.

\subsection{First Step Toward the Proof}

We begin with $\maxprodcspstar(\cdot)$'s that can be solved in polynomial time.  

\begin{proposition}\label{PO-computable}
If either $\FF\subseteq \AF$ or $\FF\subseteq \ED$, then $\maxprodcspstar(\FF)$ belongs to $\po$.
\end{proposition}

\begin{proofsketch}
For every target problem, as in the proof of \cite[Lemma 6.1]{Yam10a}, we can greatly simplify the structure of each input instance so that 
it depends only on polynomially many solutions. By examining all such solutions deterministically, we surely find its optimal solution. Hence, the target problem belongs to $\po$. 
\end{proofsketch}

It thus remains to deal with only the case where $\FF\nsubseteq\AF$ and $\FF\nsubseteq\ED$. In this case, we first make the following key claim that leads to the main theorem. 

\begin{proposition}\label{key-proposition}
Let $f$ be any constraint and assume that $f\not\in\AF\cup\ED$. Let $\FF$ be any set of constraints. 
\begin{enumerate}\vs{-2}
\item If $f\in \IMopt$, then $\maxprodcspstar(Implies,\FF)$ is APT-reducible to $\maxprodcspstar(f,\FF)$.
\vs{-2}
\item If $f\not\in \IMopt$, then there exists a constraint $g\in\{OR,NAND\}$ such that $\maxprodcspstar(g,\FF)$ is APT-reducible to $\maxprodcspstar(f,\FF)$. 
\end{enumerate}
\end{proposition}

We postpone the proof of the above proposition and, meanwhile, we want to  
prove Theorem \ref{main-theorem} using the proposition.

\begin{proofof}{Theorem \ref{main-theorem}}
If $\FF\subseteq\AF$ or $\FF\subseteq\ED$, then 
Proposition \ref{PO-computable} implies that $\maxprodcspstar(\FF)$ 
belongs to $\po$. 
Henceforth, we assume that $\FF\nsubseteq \AF$ and $\FF\nsubseteq \ED$. 
If $\FF\subseteq\IMopt$, then  
Lemma \ref{IMopt-to-FLOW} helps APT-reduce $\maxprodcspstar(\FF)$ 
to $\mathrm{MAX\mbox{-}PROD\mbox{-}FLOW}$. Next, we choose a constraint $f\in\FF$ for which $f\not\in \AF\cup \ED$. 
Proposition \ref{key-proposition}(1) then yields an APT-reduction  from  $\maxprodcspstar(Implies)$ to $\maxprodcspstar(f)$. 
By Lemma \ref{OR-complete}(2), we obtain   $\mathrm{MAX\mbox{-}PROD\mbox{-}BIS}  \APTreduces \maxprodcspstar(Implies)$. Since $\maxprodcspstar(f)\APTreduces \maxprodcspstar(\FF)$, it follows that $\mathrm{MAX\mbox{-}PROD\mbox{-}BIS}$ is APT-reducible to $\maxprodcspstar(\FF)$.

Finally, we assume that $\FF\nsubseteq \IMopt$. 
Take a constraint $f\in\FF$ 
satisfying that $f\not\in \AF\cup\ED\cup\IMopt$. In this case, 
Proposition \ref{key-proposition}(2) yields an APT-reduction from  
$\maxprodcspstar(OR)$ to $\maxprodcspstar(f)$, since $\maxprodcspstar(OR)\APTequiv \maxprodcspstar(NAND)$. {}From $\maxprodcspstar(f)\APTreduces \maxprodcspstar(\FF)$, it immediately 
follows that 
$\maxprodcspstar(OR)$ is APT-reducible to $\maxprodcspstar(\FF)$. By 
Lemma \ref{OR-complete}(1),  $\mathrm{MAX\mbox{-}PROD\mbox{-}IS} \APTreduces \maxprodcspstar(OR)$.   Therefore, we conclude that  $\mathrm{MAX\mbox{-}PROD\mbox{-}IS}$ is APT-reducible to $\maxprodcspstar(\FF)$. 
\end{proofof}

\subsection{Second Step Toward the Proof}

To finish the proof of Theorem \ref{main-theorem}, we still need to prove  Proposition \ref{key-proposition}. Proving this proposition requires three  properties. To describe them, we first review two existing notions from \cite{Yam10a}. 
We say that a constraint $f$ has {\em affine support} if $R_f$ is an affine relation and that $f$ has {\em imp support} if $R_f$ is logically equivalent to a conjunction of a certain ``positive'' number of relations of the form $\Delta_0(x)$, $\Delta_1(x)$, and $Implies(x,y)$. The notation $AFFINE$  denotes the set of all affine  relations. 

In the following three statements, $\FF$ denotes an arbitrary set of constraints.

\begin{lemma}\label{IMopt-DG-Implies}
If $f$ is a non-degenerate constraint in $\IMopt$ and has no imp support, then $\maxprodcspstar(Implies,\FF)\APTreduces \maxprodcspstar(f,\FF)$. 
\end{lemma}

\begin{proposition}\label{no-affine-and-IM2}
Let $f$ be any constraint having imp support. 
If either $f$ has no affine support or $f\not\in\ED$, then 
$\maxprodcspstar(Implies,\FF)$ is APT-reducible to $\maxprodcspstar(f,\FF)$. 
\end{proposition}

\begin{proposition}\label{no-affine-and-no-IM2}
Let $f\not\in\NZ$ be any constraint. If $f$ has neither affine support nor imp support, 
then there exists a constraint $g\in\{OR,NAND\}$ such that  $\maxprodcspstar(g,\FF)\APTreduces \maxprodcspstar(f,\FF)$. 
\end{proposition}

With a help of the above statements, we can prove Proposition \ref{key-proposition} as follows.  

\begin{proofsketchof}{Proposition \ref{key-proposition}}
Let $f\not\in\AF\cup\ED$ be any constraint. 
We proceed our proof by induction on 
the arity $k$ of $f$.  For 
the proposition's claims, (1) and (2), the basis case $k=1$ is trivial since $\ED$ contains all unary constraints. 
Next, we prove the induction step $k\geq3$. 
In the remainder of this proof,  as our induction hypothesis, 
we assume that the proposition holds for any arity less than $k$. The claims (1) and (2) will be shown separately.

(1) Assume that $f$ is in $\IMopt$. If $f$ has imp support, since $f\not\in\ED$, we can apply Proposition \ref{no-affine-and-IM2} and immediately obtain the desired APT-reduction  $\maxprodcspstar(Implies,\FF)\APTreduces \maxprodcspstar(f,\FF)$. 
Otherwise, by Lemma \ref{IMopt-DG-Implies}, we have the desired APT-reduction. 

(2) Since $f$ has no imp support, 
if $R_f$ is not affine, then Proposition \ref{no-affine-and-no-IM2} implies that, for a certain $g_0\in\{OR,NAND\}$, 
$\maxprodcspstar(g_0,\FF)\APTreduces \maxprodcspstar(f,\FF)$; therefore, the desired result follows. To finish the proof, we hereafter assume the affine property of $R_f$.  

[Case: $f\in \NZ$] 
Recall that $f\not\in\ED$ and $R_f\in AFFINE$.  
Since $f\in \NZ$, we have $|R_f|=2^k$, and thus $f$ should 
be in {\em clean form} (\ie $f$ contains no factor of the form: $\Delta_0(x)$, $\Delta_1(x)$, and $EQ(x,y)$). 
As shown in \cite[Lemma 7.5]{Yam10a}, there exists a constraint 
$p=(1,x,y,z)\not\in\ED$ with $xyz\neq0$, $z\neq xy$, and $p\leq_{con}^{max}f$. 
When $z<xy$, we can prove that, for a certain $g_0\in\{OR,NAND\}$, 
$\maxprodcspstar(g_0,\FF)\APTreduces \maxprodcspstar(p,\FF)$. 
In the case where $z>xy$, Lemma \ref{xw-yz-member-IMopt} implies 
$p\in \IMopt$. Since $p$ is obtained from $f$ by pinning operations only, we conclude that $f\in\IMopt$, a contradiction. This finishes the induction step. 

[Case: $f\not\in\NZ$]  
We first claim that $k\geq3$. Assume otherwise that $k=2$. 
Since $R_f\in AFFINE$, it is possible to write $f$ in the 
form $f(x_1,x_2) = \xi_{A}(x_1,x_2)g(x_1)$ 
after appropriately permuting variable indices. 
This places $f$ within $\AF$, a contradiction against the choice of $f$. 
Hence, $k\geq3$ holds. Moreover, we can prove the existence of 
a constraint $g\not\in\AF$ of arity $m$ for which $2\leq m<k$, 
$g\leq_{con}^{max}f$,   
and either $g\in \NZ$ or $R_g\not\in AFFINE$.  
If we can show that (*) there exists a constraint $g_0\in\{OR,NAND\}$ 
satisfying   
$\maxprodcspstar(g_0,\FF) \APTreduces \maxprodcspstar(g,\FF)$, then 
the proposition immediately follows from $g\leq_{con}^{max}f$. 
The claim (*) is split into two cases: (i) $R_g\in AFFINE$ and (ii) $R_g\not\in AFFINE$. For (i), we apply the induction hypothesis.  
For (ii), we apply Propositions \ref{no-affine-and-IM2}--\ref{no-affine-and-no-IM2}. Thus, we have completed the induction step. 
\end{proofsketchof}

In this end, we have completed the proof of the main theorem. The detailed proofs omitted in this extended abstract will be published shortly. We hope that our systematic treatment of $\maxprodcspstar$s would lead to a study on a far wider class of optimization problems.  

\bibliographystyle{alpha}

\end{document}